\documentclass[12pt]{iopart}
\usepackage{iopams}
\usepackage{enumerate}
\usepackage{amsfonts}
\usepackage{amssymb}
\usepackage{dcolumn}
\usepackage{bm}
\usepackage{graphicx, graphics}
\usepackage{color}
\usepackage[pdftex]{hyperref}
\hypersetup{
    colorlinks=true,
    linkcolor=[rgb]{0.0,0.1,0.8},
    citecolor=[rgb]{1.0,0.0,0.0},
    filecolor=magenta,
    urlcolor=[rgb]{0.0,0.3,1}
}

\newcommand{\pac}[1]{ \left\{ #1 \right\} }
\newcommand{\pap}[1]{\left( #1 \right)}
\newcommand{\pas}[1]{\left[#1 \right]}

\newcommand{\sio}{{\sigma}}

\newcommand{\Jo}{{J}}

\newcommand{\bra}[1]{\left\langle #1 \right\vert}

\newcommand{\ket}[1]{\left\vert #1 \right\rangle}
\newcommand{\lam}{\lambda}
\newcommand{\ome}{\omega}
\newcommand{\eps}{\epsilon}

\begin{document}
\title[Dynamics of Entanglement and the Schmidt Gap in a Driven Light-Matter System]{Dynamics of Entanglement and the Schmidt Gap in a Driven Light-Matter System}

\author{F.~J. G\'omez-Ruiz$^1$,
J.~J. Mendoza-Arenas$^{1,2}$,
O.~L. Acevedo$^3$,
F.~J. Rodr\'iguez$^1$,
L. Quiroga$^1$
and
N.~F. Johnson$^4$}

\address{$^1$ Departamento de F{\'i}sica, Universidad de los Andes, A.A. 4976, Bogot{\'a} D. C., Colombia}
\address{$^2$ Clarendon Laboratory, University of Oxford, Parks Road, Oxford OX1 3PU, United Kingdom}
\address{$^3$ JILA, University of Colorado, Boulder, CO 80309, U.S.A.}
\address{$^4$ Department of Physics,  University of Miami, Coral Gables, FL 33124, U.S.A.}
\ead{\href{mailto:fj.gomez34@uniandes.edu.co}{fj.gomez34@uniandes.edu.co}}

\begin{abstract}
 {\bf The ability to modify light-matter coupling in time (e.g. using external pulses) opens up the exciting possibility of generating and probing new aspects of quantum correlations in many-body light-matter systems. Here we study the impact of such a pulsed coupling on the light-matter entanglement in the Dicke model as well as the respective subsystem quantum dynamics. Our dynamical many-body analysis exploits the natural partition between the radiation and matter degrees of freedom, allowing us to explore time-dependent intra-subsystem quantum correlations by means of squeezing parameters, and the inter-subsystem Schmidt gap for different pulse duration (i.e. ramping velocity) regimes -- from the near adiabatic to the sudden quench limits. Our results reveal that both types of quantities indicate the emergence of the superradiant phase when crossing the quantum critical point. In addition, at the end of the pulse light and matter remain entangled even though they become uncoupled, which could be exploited to generate entangled states in non-interacting systems.}
\end{abstract}
\noindent{\it Keywords\/}: Non-Equilibrium Dicke Model, Schmidt Gap, Entanglement, Squeezing.\\
\\
\submitto{\jpb}
\maketitle

\section{Introduction}
The understanding, characterization and manipulation of non-equilibrium correlated many-body systems has benefitted from several remarkable experimental and theoretical advances in recent years~\cite{Nori_RMP2014, Zoller_NJP2011}. Although, by definition, any laboratory sample will necessarily interact with its laboratory environment~\cite{breuer}, modern technologies have succeeded in isolating quantum systems to a significant degree within a large variety of experimental settings~\cite{Lloyd_SC96, Schneider_RP2012, Houck_Nat2012}. Many of these realizations can be regarded as particular cases of an interaction between matter and radiation, or some other form of bosonic excitation field. 
From a theoretical point of view, many of these systems can be modeled to a reasonable approximation by considering the matter subsystem as two-level systems (qubits) and the radiation subsystem as a set of independent harmonic oscillators. Examples of such modeling include cavity Quantum Electrodynamics (QED)~\cite{nori2011nature,xiang2013rmp} and circuit QED~\cite{niemczyk2010nature,peterson2012nature}, impurities immersed in Bose-Einstein condensates (BECs)~\cite{ng2008pra,haikka2011pra,sabin2014scirep}, and artificial atoms of semiconductor heterostructures interacting with light~\cite{Sarah_IOP2012} or with plasmonic excitations~\cite{Dzsotjan2010prb}. Since these systems contain various degrees of freedom, their theoretical study has been traditionally approached using approximate perturbative methods~\cite{breuer}. 

Most of the theoretical treatments to date rely on the assumption that the matter-radiation interaction is static, and either very weak or very strong. However from an empirical perspective, these regimes do not represent any technological boundary -- indeed, the coupling strength in real systems is quite likely to be in between these limits. The potential richness of effects in this intermediate case {\em and} in the regime of non-static coupling, is therefore of significant interest for temporal quantum control in practical quantum information processing and quantum computation. On a more fundamental level, an open-dynamics quantum simulator would be invaluable for shedding new light on core issues at the foundations of physics, ranging from the quantum-to-classical transition and quantum measurement theory~\cite{Zurek1} to the characterization of Markovian and non-Markovian systems~\cite{Bruer_PRL, PRBluis,Cosco2017arxiv}.

In our work we explore this dynamical regime which is opened up by manipulating the strength of the light-matter coupling in time -- for example using external pulses that generate a coupling that cycles from weak to strong and back again. Specifically, we use a general, time-dependent many-body Hamiltonian, namely the Dicke model, to study the impact of a single pulse in the light-matter coupling, on the quantum correlations at the collective and subsystem levels. Exploiting the natural partition between the radiation and matter degrees of freedom, we explore the time-dependent squeezing parameters of each subsystem, and the entanglement spectrum through the Schmidt gap, for different pulse duration (i.e. ramping velocity) regimes, ranging from the near-adiabatic to the sudden quench limits. The results show that both the inter-subsystem {\em and} and intra-subsystem quantum correlations signal the emergence of the superradiant phase. In addition, in the intermediate ramping regimen, both subsystems remain entangled at the end of the applied pulse, which should be of interest for quantum control schemes.

The paper is organized as follows. In Section~\ref{sect_DM} we describe the time-dependent light-matter model that we analyze. In Section~\ref{sect_results} we present our main results for the driven dynamics, starting with the coherence and squeezing of the light and matter subsystems, and then considering the Schmidt gap for the light-matter bipartition. We provide a discussion of our results in Section~\ref{sect_discussion}. Finally we present our conclusions in Section~\ref{sect_conclu}.

\section{Model and methods}\label{sect_DM}
The time-dependent light-matter system that we choose for our study is the non-equilibrium Dicke Model (NE-DM)~\cite{Acevedo2014PRL}, which consists of a set of qubits coupled to a single radiation mode. Its Hamiltonian is given by $\left(\hbar=1\right)$
\begin{equation}\label{Hdicke}
\hat{H}\pap{\lam\pap{t}}=\epsilon \hat{J}_{z} + \omega \hat{a}^{\dagger}\hat{a} +\frac{2\lambda(t)}{\sqrt{N}}\hat{J}_{x}\left(\hat{a}^{\dagger}+\hat{a}\right).
\end{equation}
Here $N$ is the number of qubits, $\hat{J}_{\alpha}=\frac{1}{2}\sum_{i=1}^{N}\sio_{\alpha}^{\pap{i}}$ $\left(\alpha=x,z\right)$ denote collective angular momentum operators of the qubits acting on the totally symmetric manifold (known as the {\em Dicke manifold}), $\hat{a}^{\dag} \pap{\hat{a}}$ is the creation (annihilation) operator of the radiation field, $\epsilon$ and $\omega$ represent the qubit and field transition frequencies respectively, and $\lambda\pap{t}$ is the time-dependent coupling between matter and light subsystems. For all the numerical results in this paper, we consider the resonant case between the qubits and the radiation frequency, and set the energy scale by taking $\epsilon=\omega=1$. Though we focus on light-matter systems, other realizations are possible -- including those where the bosonic mode corresponds to vibrational degrees of freedom. 

The static properties of Dicke Model have been widely studied and characterized in the last two decades~\cite{Nagy2010prl,Das2016njp}. It is well known that, in the thermodynamic limit $N\to\infty$, it exhibits a second-order quantum phase transition (QPT) \cite{Hioe_PRA1973} at $\lam_{c}=\sqrt{\eps\ome}/2$ with order parameter $\hat{a}^{\dag}\hat{a}/J$, separating the normal phase at $\lam_{c}<\sqrt{\eps\ome}/2$ from the superradiant phase in which there is a finite value of the macroscopic order parameter, e.g. finite boson expectation number \cite{Acevedo2014PRL}. Now we discuss its dynamical properties under the time-dependent model in Eq.~\ref{Hdicke}. We obtain the full NE-DM instantaneous state $\ket{\psi\pap{t}}$ by numerically solving the time-dependent Schr\"odinger equation. Our numerical solution of the NE-DM profits from the fact that the operator $\hat{{\bf J}}^{2}=\sum_{\alpha}\hat{J}_{\alpha}^{2}$ is a constant of motion with eigenvalue $J\left(J+1\right)$, and that the parity operator $\hat{\mathcal{P}}=\exp\left(\imath \pi\left[\hat{a}^{\dag}\hat{a} +\Jo_{z}+J\right]\right)$ is also conserved and commutes with $\hat{{\bf J}}^{2}$. Since we are seeking results that have general validity, we avoid making the rotating-wave approximation that is commonly used to solve the static version of the NE-DM and which makes it Bethe ansatz integrable~\cite{Gaudin_JPF1976}. The general structure of the state $\ket{\psi\pap{t}}$ at any time $t$ is given by
\begin{equation}\label{state}
\ket{\psi\pap{t}}=\sum_{m_{z}=-N/2}^{N/2}\sum_{n=0}^{\chi}C_{n,m_{z}}\pap{t}\ket{m_{z},n}.
\end{equation}
Here $\chi$ is the truncation parameter of the size of the bosonic Fock space, whose value we choose to be large enough to ensure that the numerical results converge  \cite{Acevedo2014PRL}. The basis states $\ket{m_{z},n}=\ket{m_{z}}\otimes\ket{n}$ are defined such that $\ket{m_{z}}$ is an eigenvector of $J_{z}$ in the subspace of even parity with eigenvalue $m_z$, and $\ket{n}$ is a bosonic Fock state with occupation $n$. The initial state of the dynamics at $t=0$, with negligible light-matter coupling $\lambda(t)=0$, is the non-interacting ground state $\ket{\psi(0)}=\bigotimes_{i=1}^{N}\ket{\downarrow}\otimes\ket{n=0}=\ket{-\frac{N}{2},0}$, where both the matter and light subsystems have zero excitations. All qubits are polarized in the state with $\langle\sigma_z\rangle=-1$, and the field is in the Fock state of zero photons. Since the total angular momentum and parity are conserved quantities, we can without loss of generality restrict our study to the maximum angular momentum sector $J=N/2$ and $\mathcal{P}=1$.

We obtain the time-evolution of the system in response to an up-down pulse in $\lambda(t)$, which is chosen so that the system dynamically crosses the QPT on both the up and down portion of the $\lambda(t)$ pulse cycle. For simplicity and considering the successful and actual experimental platform for driving light-matter interaction in Dicke Model~\cite{ExpDicke1, ExpDicke2, ExpDicke3}, we consider $\lambda(t)$ to rise and fall linearly in time $t$ during the pulse, i.e. it has the form $\upsilon t$, and thus establishes a triangular ramping of the light-matter interaction. The rate $\upsilon=\frac{d \lambda}{d t}$ acts as a control parameter and is known as the {\it annealing velocity}, which is characterized by a finite time $\tau$ such that $\upsilon=1/\tau$. The particular choice of $\lambda(t)$, implies that the quantum critical point is crossed twice during the cycle, first when $t = \tau/2$, and second when $t = 3\tau/2.$

\begin{figure}[t]
\begin{center}
\includegraphics[scale=0.7]{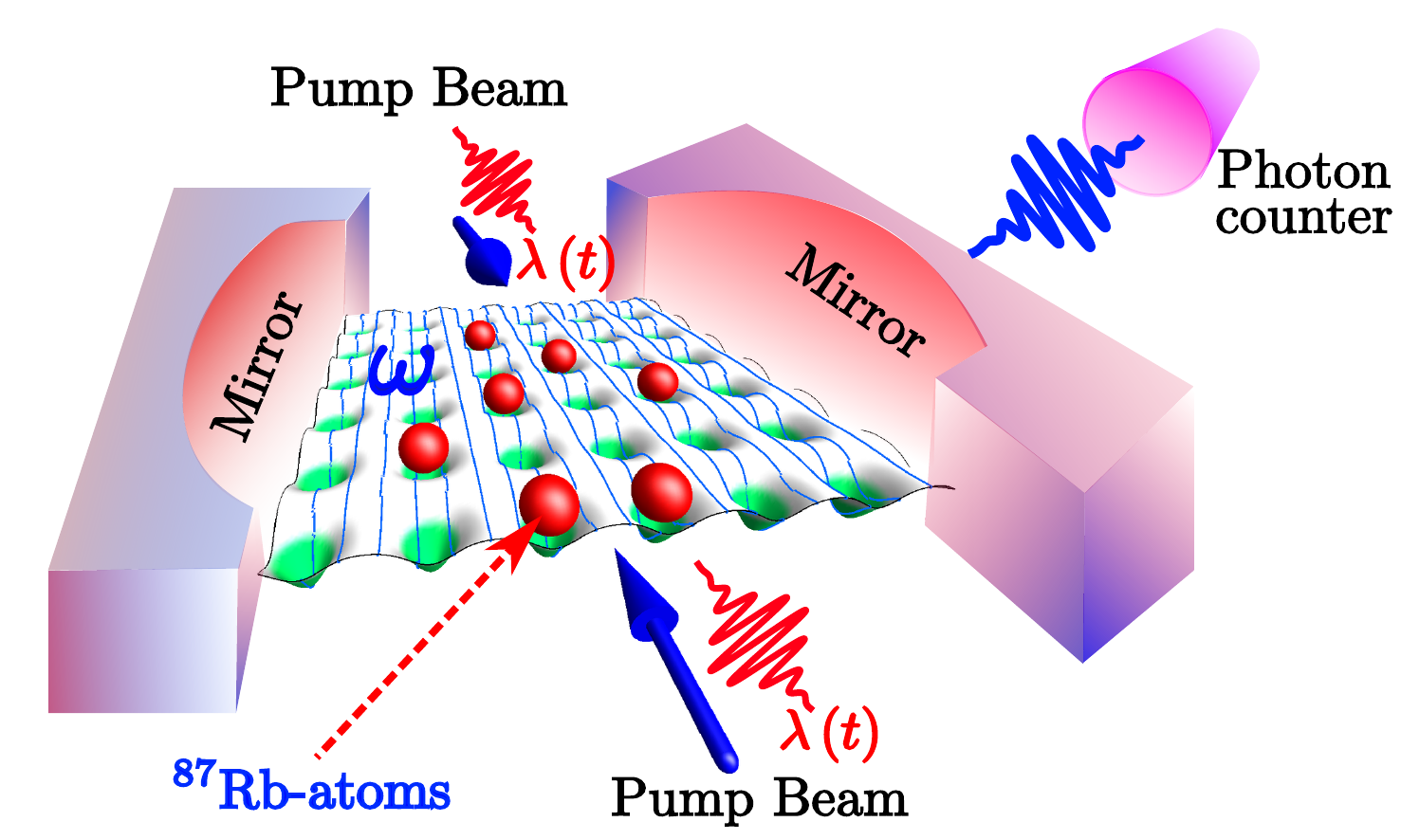}
\caption{(Color online) Schematic of recent successful realizations of the Dicke Model in a BEC of $^{87}$Rb atoms confined by a magneto-optical trap. The atoms (qubits) are made to interact with the  field mode of a high  finesse cavity by means of a pumped transverse field.~\cite{Klinder_PNA2015} }\label{fig_1}
\end{center}
\end{figure}

Several experimental realizations of the DM have been discussed during the past few  decades, with many being built around implementations in circuit QED where superconducting qubits play the role of the matter subsystem~\cite{ExpDicke1, ExpDicke2, ExpDicke3}. Additionally to date, important experimental scenarios have shown efficient and effective ways to simulate radiation-matter interaction systems with time-varying couplings. Some of the most promising experimental possibilities are thermal gases of atoms~\cite{BlackPLR}, and BECs using momentum~\cite{ExpDicke4} and hyperfine states~\cite{BaumannPRL}. However, we believe that the branch of experiments deserving most attention is that demonstrating DM superradiance in ultra-cold atom optical traps, especially $^{87}$Rb Bose-Einstein condensates~\cite{Klinder_PNA2015, ExpDicke4, ExpDicke5, ExpDicke6}. 
Indeed, these atom-trap experiments are so promising that a brief description of them is pertinent here, following Ref.~\cite{Klinder_PNA2015}. Figure~\ref{fig_1} depicts a schematic of the main components of the atom-trap DM realization. It corresponds to an ultracold cloud of $N\sim 10^{5}$ $^{87}$Rb atoms confined by a magneto-optical trap inside a high  finesse Fabri-Perot cavity. The cloud is driven by a transverse pump laser whose wavelength is the same as that of the fundamental mode of the cavity. The combined cavity and pump laser setting produces an optical lattice potential that affects the motion of the atoms in the cloud through coupling with far-detuned atomic resonances. This coupling causes the atoms to interact with each other through the mediating presence of the radiation mode. At low intensity pump power $\epsilon_{p}$, the BEC remains in its (almost spatially uniform) translational ground state. However, when a critical value of $\epsilon_p$ is reached, the ground state becomes a grid-like matter wave as the one shown in Fig~\ref{fig_1}. This change of configuration constitutes the QPT whose spontaneous symmetry breaking is caused by the fact that two matter wave configurations, which are distinguishable only by a phase difference of $\pi$, have the same lowest possible energy. The effective two-level (qubit) system is composed of the ground BEC translational state and the fundamental grid-like matter wave state for each atom. There are several ways to monitor the system. The two most fundamental are (1) addressing the radiation  field by coupling one of the (unavoidably) leaky walls of the cavity to a detector, and (2) using time-of-flight methods to measure the matter wave modes. Note that each technique measures the state of one of the two main components of the DM, i.e. light and matter respectively. 

\section{Results}\label{sect_results}
We now proceed to characterize the complete dynamical QPT profile by focusing on properties of each subsystem, namely the matter subsystem composed by the all-to-all (qubit) spin network, and the radiation mode subsystem. We analyze a wide range of annealing velocities $v$, and use a logarithmic scale for showing these values of the velocity, defined by $\Gamma=\log_2(v)$. This range varies from the slow near adiabatic regime,  through the intermediate regime, to the fast sudden-quench regime. In previous papers, we showed how the values of the velocity for which the change of regime is manifested depend on system size~\cite{Gomez_Ent2016,Acevedo2014PRL, Acevedo2015NJP, AcevedoPRA2015}. Here, on the other hand, we take a fixed size of the qubit subsystem, namely $N=81$, for all the numerical calculations. 

\subsection{Coherence and Squeezing}
\begin{figure}[t]
\begin{center}
\includegraphics[scale=0.8]{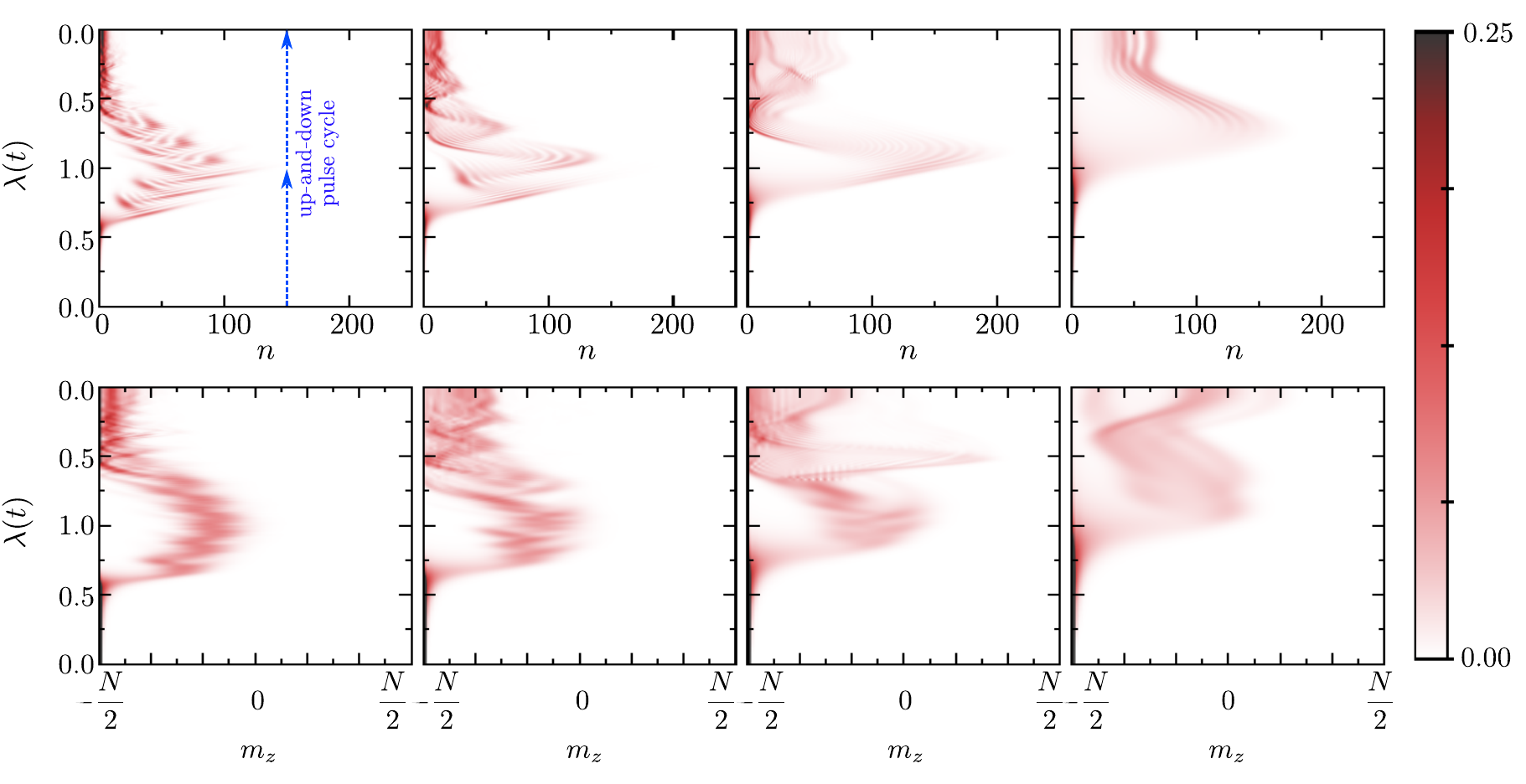}
\caption{(Color online) Projection of reduced density matrix for each subsystem. Upper panels: $\langle n | \rho_{B}| n\rangle$. Lower panels: $\langle m_z | \rho_{Q}| m_z\rangle$. The values of velocities are, from left to right: $\Gamma=-7.76,\, -5.76,\, -4.76,\, -2.76$. The color scale was adjusted to improve visualization of the results.}\label{fig_2}
\end{center}
\end{figure}

We begin our discussion of the driven dynamics of the Dicke model by considering the diagonal elements of the reduced density matrices of the bipartition. Figure~\ref{fig_2} shows the instantaneous projection of the reduced density matrix for the matter and radiation subsystems over the $J_z$ and Fock basis respectively, for several values of the annealing velocity. For the slowest ramping ($\Gamma=-7.76$), both the radiation mode and qubits remain entirely unexcited before crossing the quantum critical point $\lambda_c$ when increasing $\lambda$, with only the respective states of $n=0$ and $m_z=-N/2$ being populated. After crossing $\lambda_c$ the population is transferred to states of larger $n$ and $m_z$, a process that continues up to the time where $\lambda$ starts decreasing. With the reversal of $\lambda$ the population of large values of $m_z$ and $n$ is transferred back to lower values, in a highly-symmetrical form with respect to the turning point. When $\lambda_c$ is crossed again, both the radiation field and the set of qubits become almost completely unexcited, with only the lowest values of $m_z$ and $n$ being populated. Since the reversal of the matter and light dynamics is not completely achieved, this corresponds to a near-adiabatic regime instead of a true adiabatic one (e.g. see results on Section~\ref{gap_section} for lower velocities).    
For larger annealing velocities, the dynamic population of states with $m_z>-N/2$ and $n>0$ remains qualitatively similar to that of the near-adiabatic limit during the linear increase of $\lambda$. However two main qualitative differences are observed. First, this population transfer occurs further and further away from $\lambda_c$ as $v$ increases, indicating that the ground-state QPT is not being immediately captured. Second, larger values of $n$ and $m_z$ are reached, since a faster ramping velocity provides a stronger excitation to the system. On the other hand, the population dynamics of the $\lambda$ reversal regime is very different to that close to adiabaticity. Even though the population is also transferred back to states of lower quantum numbers, the symmetry with respect to the turning point is lost, and at the end of the dynamics, when the matter and radiation become uncoupled, they are still highly excited. This already indicates that for large annealing velocities, the system gets so excited that it does not simply follow the decrease of $\lambda$, which is of course only expected in the adiabatic limit. Similar asymmetric results are found in the squeezing and entanglement spectrum results shown below.   

\begin{figure}[t]
\begin{center}
\includegraphics[scale=0.8]{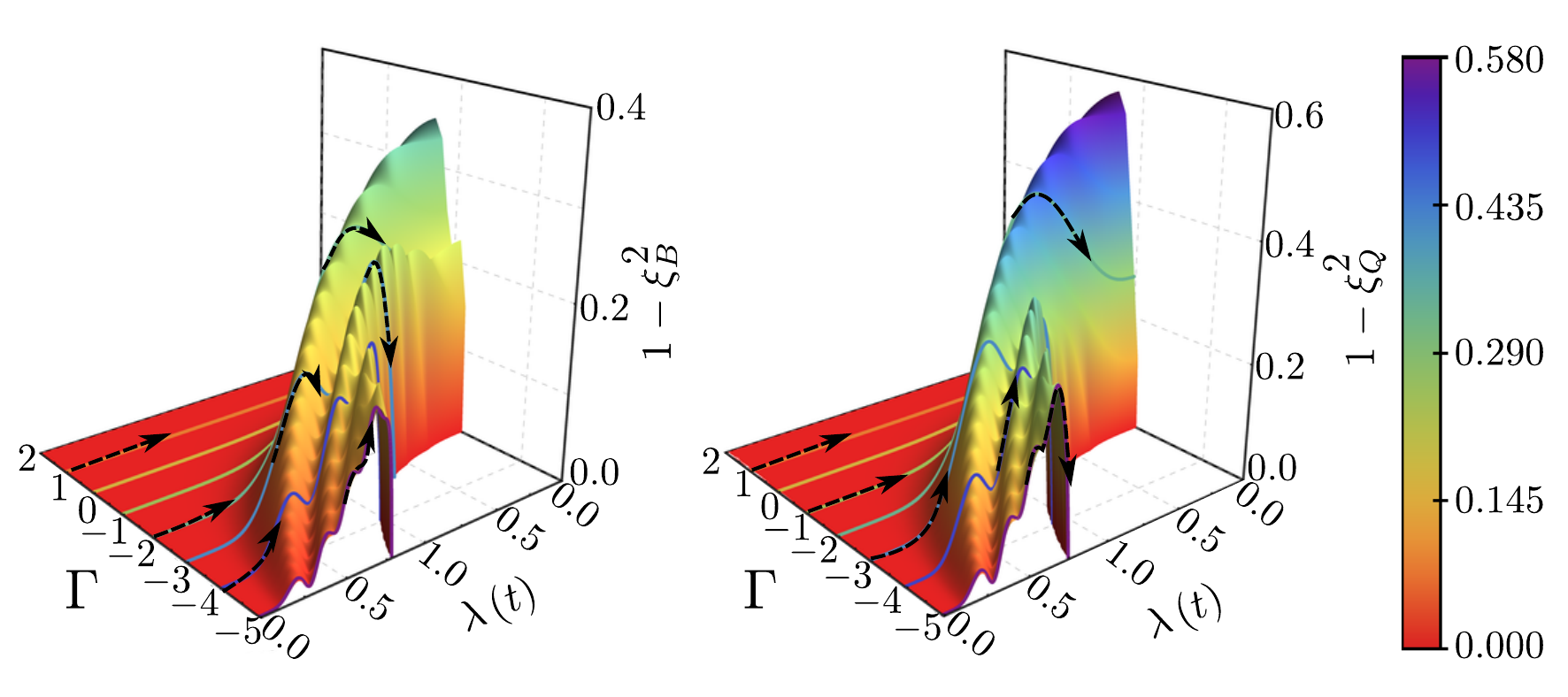}
\caption{(Color online) Dynamic profiles of the matter subsystem in which, for fixed annealing velocities, time varies according to the direction of arrows. Left panel: Evolution $1-\xi_{B}^{2}$, as defined in Eq.~\ref{Sbos}, whenever it is greater that zero (squeezed radiation). Right: Two qubit concurrence $c_{w}\pap{N-1}=1-\xi_{Q}^{2}$. }\label{fig_3}
\end{center}
\end{figure}

Now we describe the squeezing parameter for both subsystems, starting with the light degrees of freedom. The squeezing of light states has widely been studied in the literature. A squeezed state of light arises in a simple quantum model comprising non-linear optical processes such as optical parametric oscillation and four-wave mixing. The fundamental importance of the squeezed state is characterized by the property that the variance of the quadrature operator $\hat{x}$ is less than the value $1/2$ associated with the vacuum and coherent state. The squeezing parameter in the field mode $\xi_{B}^{2}$ is expressed in terms of the variance (Var) and covariance (Cov) of the field quadratures as~\cite{Walls}              
 \begin{equation}\label{Sbos}
  \xi_{B}^{2} = {\rm Var}\pap{\hat{x}} + {\rm Var}\pap{\hat{p}} - \sqrt{\pap{{\rm Var}\pap{\hat{x}}-{\rm Var}\pap{\hat{p}}}^2 + 4 {\rm Cov}\pap{\hat{x},\hat{p}}^2}.
 \end{equation} 
In the left panel of Figure~\ref{fig_3}, we present a novel way to generate a photon squeezed state. At $t=0$ the quantum cavity starts in the vacuum state $|n=0\rangle$. As before, the radiation-matter parameter varies as a simple linear up-down pulse, forming a triangular ramping. Our results show the existence of a specific regime of annealing velocities such that while the pulse is applied, the photon squeezing tends to increase (besides small oscillations) even after the reversal ramping of $\lambda(t)$ has started. Furthermore, we note that for this velocity regime, the final state of light has high squeezing when the final radiation-matter parameter is zero. 
                
Now we discuss the dynamics of spin squeezing, which has also been the object of intense research in the past few decades. For example, the natural idea of transferring squeezing from light to atoms has been attracting attention both theoretically and experimentally. The notion of spin squeezing has arisen mainly from two considerations: the study of particle correlations and entanglement~\cite{Wang_PRA2003, Zoller, Nick}, as well as the improvement of measurement precision in experiments~\cite{Wineland}. The experimental proposals for transferring squeezing from light to atoms include placing the latter in a high-Q cavity so they interact repeatedly with a single-field (not squeezed) mode~\cite{Ueda}, and illuminating bichromatic light on atoms in a bad cavity~\cite{Klaus}. The intrinsic spin squeezing in a large atomic radiating system was studied in Ref.~\cite{yukalov}, where spin-squeezed states were generated by means of strong interatomic correlations induced by photon-exchange. Spin squeezing can also be produced via a squeezing exchange between motional and internal degrees of freedom of atoms~\cite{Saito}.  For a detailed review, we refer to Ref.~\cite{Jian}.  

The definition of spin-squeezing is not unique~\cite{Jian}. For our propose we use the definition given in Ref.~\cite{Wang_PRA2003}, in which a relation between entanglement for a two-qubit subsystem as measured by the Wootters concurrence $c_{w}$~\cite{wootters} and the spin squeezing parameter $\xi_{Q}$ was established, namely
\begin{equation} \label{squee_concu}
\xi_{Q}^{2}=1-\pap{N-1}c_{w}.
\end{equation}
Since each qubit is equally entangled with each other, the monogamic character of entanglement is manifested in Eq.~\ref{squee_concu} by the $N-1$ factor.

In the right panel of Fig.~\ref{fig_3} we show the spin squeezing for a wide range of velocities. We find a regime of intermediate annealing velocities for which the squeezing is large at the end of the pulse, which coincides with the velocity regime for which the photonic squeezing is magnified. In previous works by some of us, we showed that the intermediate velocity regime allows for the generation of entanglement~\cite{Gomez_Ent2016,Acevedo2015NJP}; this is manifested in the generation of squeezing in both light and matter. A fundamental and novel feature of our results is that there is no need of ultra-strong coupling to have squeezing in both light and matter. In addition, we note that the squeezing after the pulse widely exceeds the values that would be achieved through a near-adiabatic evolution.     

\subsection{Schmidt gap} \label{gap_section}
The observation several years ago of the fundamental role of entanglement on quantum criticality led to intense research on characterizing QPTs by means of different measures such as entanglement entropy and concurrence~\cite{amico2002nature,gu2003pra,latorre2003prl,wu2004prl,laflorencie2006prl,zanardi2006njp,buonsante2007prl,amico2008rmp,jjma2010pra,pino2012pra,hofmann2014prb}. Shortly after, it was shown that the entanglement spectrum, i.e. the set of eigenvalues of the reduced density matrix of one subsystem resulting from a bipartition, provides valuable information on the properties of topological phases~\cite{haldane2008prl}, and remarkably even more than the entanglement entropy. Since then, several works have analyzed the behavior of the entanglement spectrum, and in particular of the Schmidt gap (the difference between the two largest eigenvalues) close to criticality for different scenarios. These include zero-temperature QPTs~\cite{chiara2012prl,lepori2013prb,bayat2013nat}, where the Schmidt gap has been suggested as an order parameter, many-body localization~\cite{Gray2017arxiv}, and dynamical crossings of QPTs at different speeds~\cite{canovi2014prb,torlai2014jstat,Qijun2015prb,francuz2016prb,coulamy2017pre}. The latter situation, corresponding to our point of interest in the present work, has been mostly studied for quantum spin chains. Now we discuss the dynamics of the Schmidt gap of the non-equilibrium Dicke model.   

In contrast to several condensed-matter systems, the Dicke model immediately suggests a bipartition which allows for a direct study of the physical properties of subsystems of different nature, i.e.  the set of qubits and the radiation field. Thus we calculate the entanglement spectrum for this bipartition. In general, the dynamical state of the total system $\ket{\psi\pap{t}}$ is represented by the bipartite form in Eq.~\ref{state}. A standard singular-value-decomposition therefore allows us to rewrite this state as 
\begin{equation} \label{schmidt}
\ket{\psi\pap{t}}=\sum_{\alpha=1}^{\Xi}S_{\alpha}\pap{t}\ket{\Phi_{\alpha}^{\pas{m_{z}}}\pap{t}}\otimes \ket{\Phi_{\alpha}^{\pas{n}}\pap{t}},
\end{equation}
with
\begin{eqnarray*}
\ket{\Phi_{\alpha}^{\pas{m_{z}}}\pap{t}}&=\sum_{m=-N/2}^{N/2} U_{m,\alpha}\pap{t} \ket{m_{z}},\qquad\ket{\Phi_{\alpha}^{\pas{n}}\pap{t}}&=\sum_{n=0}^{\chi} V_{\alpha,n}\pap{t} \ket{n},
\end{eqnarray*}
and where the unitary matrices ${\bf U}$ and ${\bf V}$ are defined on the corresponding subspaces $\mathcal{H}_{m_z}$ and $\mathcal{H}_{n}$ of the set of qubits and radiation field respectively. The new orthonormal states $\pac{\ket{\Phi_{\alpha}^{\pas{m_{z}}}\pap{t}}}$ and $\pac{\ket{\Phi_{\alpha}^{\pas{n}}\pap{t}}}$ are known as Schmidt states. The diagonal elements $S_{\alpha}\geq 0$ in the expansion of Eq. \ref{schmidt} are the Schmidt coefficients, which satisfy $\sum_{\alpha}S_{\alpha}^{2}=1$ due to the normalization of the state and are assumed to be arranged in descending order with $\alpha$. Finally $\Xi=\min(N+1,\chi+1)$ is the Schmidt rank, which corresponds to the total number of coefficients in the decomposition.\\

The reduced density matrices for the two subsystems, $\rho_{m_z}\pap{t}=tr_{n}\pap{\ket{\psi\pap{t}}\bra{\psi\pap{t}}}$ and $\rho_{n}\pap{t}=tr_{m_z}\pap{\ket{\psi\pap{t}}\bra{\psi\pap{t}}}$, follow directly from the Schmidt decomposition of Eq.~\ref{schmidt} and are given by
\begin{eqnarray*}
 \rho_{m_z}\pap{t}&=\sum_{\alpha=1}^{\Xi} S_{\alpha}^{2}\pap{t}\ket{\Phi_{\alpha}^{\pas{m_{z}}}\pap{t}}\bra{\Phi_{\alpha}^{\pas{m_{z}}}\pap{t}}\\
\rho_{n}\pap{t}&=\sum_{\alpha=1}^{\Xi} S_{\alpha}^{2}\pap{t}\ket{\Phi_{\alpha}^{\pas{n}}\pap{t}}\bra{\Phi_{\alpha}^{\pas{n}}\pap{t}},
 \end{eqnarray*}
 which immediately shows that both $\rho_{m_z}\pap{t}$ and $\rho_{n}\pap{t}$ are diagonal in their respective Schmidt basis and have identical spectra. As a result, the Schmidt gap $\Delta_S$ is defined as
 \begin{equation}
 \Delta_S \equiv \left| S^2_{2}-S^2_{1}\right|,
 \end{equation}
corresponding to the difference between the two largest eigenvalues of the reduced density matrix of any of the two subsystems, and is thus a property shared by both. In the following we describe the behavior of the Schmidt gap as the QPT of the Dicke model is crossed with the triangular ramping at different annealing velocities $v$.\\

\begin{figure}[h!]
\begin{center}
\includegraphics[scale=0.8]{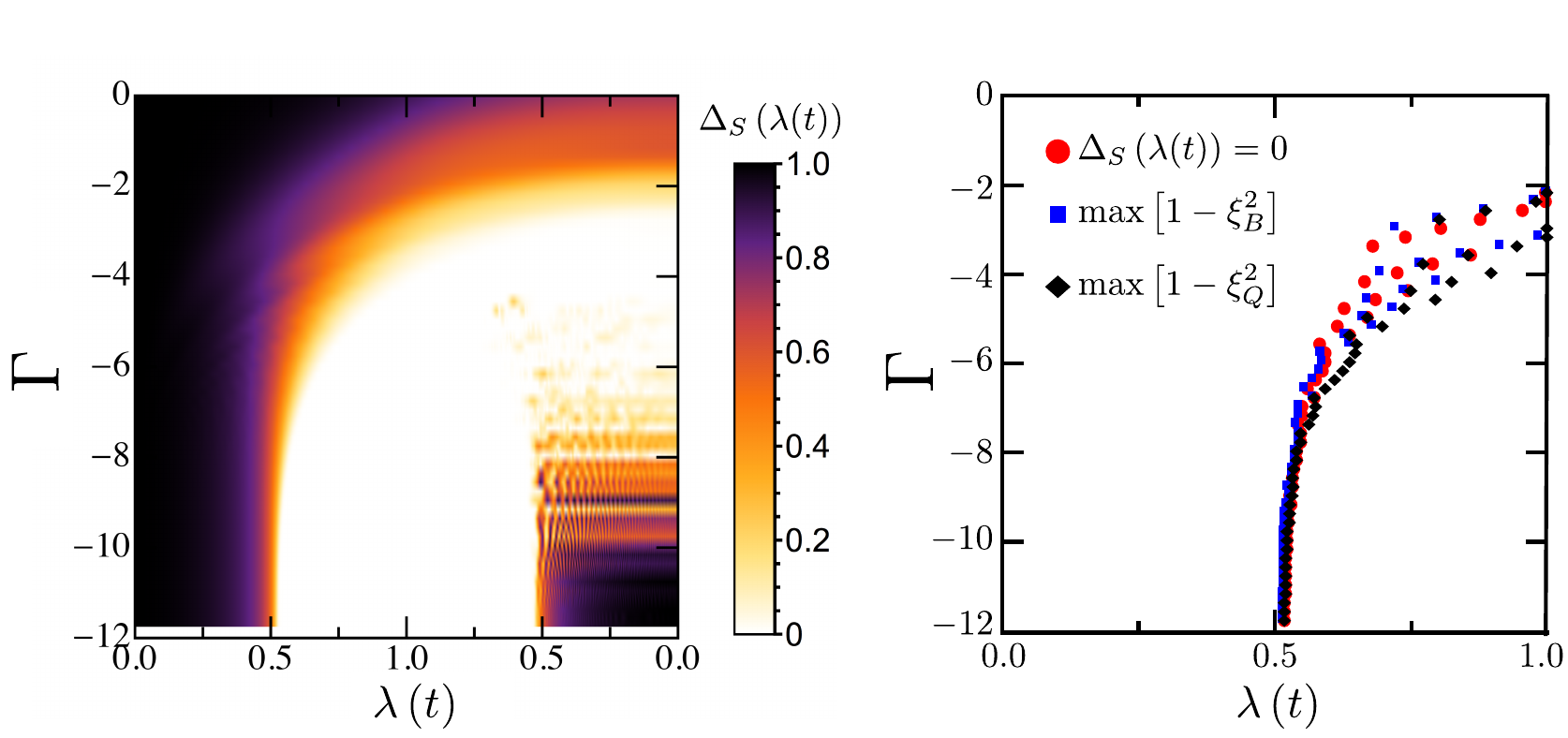}
\caption{(Color online) Left panel. Schmidt gap $\Delta_S$ as a function of the annealing velocity $\Gamma=\log_2(v)$ and the time-dependent light-matter interaction $\lambda(t)$. Right panel. Comparison of the $\lambda$ values where the Schmidt gap vanishes, and where the maximal squeezing of both qubits and photons takes place, for the same annealing velocities of the left panel and the ramping of increase of $\lambda$.}\label{fig_4}
\end{center}
\end{figure}

We first consider the crossing of the quantum critical point $\lambda_c=1/2$ during the linear increase of $\lambda(t)$, corresponding to the $0\rightarrow1$ regime in the left panel of Fig~\ref{fig_4}. Since the initial state $\ket{\psi(t=0)}$ is simply a product, only $S_1(t=0)=1$ is finite, while the other Schmidt coefficients are zero; thus $\Delta_S(t=0)=1$. During the subsequent dynamics $\Delta_S$ monotonically decreases, at a $\Gamma-$dependent rate. In the near adiabatic regime ($\Gamma\lesssim-10$) $S_1$ and $S_2$ cross and the Schmidt gap closes slightly above $\lambda_c$, which suggests that it actually captures the QPT between normal and superradiant states. This is similar to previous results of adiabatic dynamical crossings of QPTs in spin chains~\cite{canovi2014prb,torlai2014jstat,Qijun2015prb}, where the gap closes near to the corresponding transition. However, in contrast to these cases where the Schmidt coefficients separate and continue crossing during the subsequent dynamics, here $\Delta_S$ remains being zero. As the annealing velocity increases up to the intermediate regime, the Schmidt gap maintains the same qualitative decay with $\lambda$, closes further away from $\lambda_c$ similarly to dynamical crossings on spin chains, and remains zero afterwards. However for even faster ramping processes, in the sudden quench regime ($-3\lesssim\Gamma$), the decay of the gap is so slow that it remains finite when the reversal of $\lambda$ begins.

Now we discuss the dynamical crossing when $\lambda$ is reversed, depicted in the $1\rightarrow0$ regime of the left panel of Fig~\ref{fig_4}. The main feature of the near adiabatic ramping is that slightly above $\lambda_c$ the Schmidt gap becomes finite again, signaling the return of the system to the normal phase. Moreover, the system actually goes back to the initial product state $\ket{\psi(0)}$, since the Schmidt gap reaches the value $\Delta_S=1$ when $\lambda=0$. For higher annealing velocities ($-10\lesssim\Gamma\lesssim-7$) the gap shows an initial fast non-monotonic growth, after which it tends to saturate to a finite value following an oscillatory dynamics. This indicates that even though the qubit and radiation subsystems become disconnected at the end of the pulse, the total state is not just a simple product but an entangled configuration. Thus this intermediate far-from-adiabatic triangular ramping could be exploited as a protocol for preparing entangled states of non-interacting subsystems. For larger annealing velocities but before the sudden quench regime, where the Schmidt gap became zero before starting the light-matter coupling reversal ($-7\lesssim\Gamma\lesssim-5$), it emerges again before crossing $\lambda_c$ but exhibits complex  dynamics including more points of closure. For somewhat higher velocities we observe a scenario where the gap remains finite during the first stage of the driving, but since the dynamics is not so slow it still becomes zero shortly after the start of the reversal stage, before crossing $\lambda_c$ for the second time ($-5\lesssim\Gamma\lesssim-3$). This no longer occurs in the sudden quench regime, where due to the very slow dynamics the Schmidt gap never closes.

\section{Discussion} \label{sect_discussion}
The results presented in Section~\ref{sect_results}, in particular the similar qualitative profiles of the squeezing parameters and the Schmidt gap as a function of $\Gamma$ and $\lambda$, suggest that both might serve as indicators of the same non-equilibrium phenomenon. Now we briefly discuss this connection, along with a simple approach to the problem, and a possible future application. 
 
\subsection{Squeezing functions and Schmidt gap}
In the right panel of Fig.~\ref{fig_4} we show, for each annealing velocity considered, the value of $\lambda$ at which the Schmidt gap becomes zero during its increase ramping. As previously discussed, the gap vanishes at higher values of $\lambda>\lambda_c$ as the velocity increases, moving away from the near adiabatic limit. Close to the sudden quench regime ($-5\lesssim\Gamma\lesssim-2$) this general trend continuous, even though the increase is non-monotonic as the dynamics (and thus determining the exact closing point) becomes more involved. In spite of this behavior we find that remarkably, the closure of the gap coincides (quite well for low velocities, approximately for high velocities) with the points in which the maximal squeezing parameters of both qubits and photons take place. This is also shown in the right panel of Fig.~\ref{fig_4}, where the different scenarios are plotted simultaneously.

Furthermore this also agrees with the values of $\lambda$ in which the qubit and radiation order parameters become finite (see Ref.~\cite{Acevedo2015NJP}). Thus the Schmidt gap can be considered as a complementary quantity to the order parameters of the Dicke model~\cite{Qijun2015prb}, as the former is finite when the latter are zero and vice versa~\cite{Acevedo2015NJP}. These results suggest that both the Schmidt gap and the squeezing parameters are indicators of the emergence of the superradiant state when dynamically crossing the QPT, even at high velocity. 

The behavior of these quantities is far more complex during the reversal stage. Due to the strongly-oscillating behavior at low velocities and the more erratic dynamics at high velocities, determining correctly the vanishing point of the Schmidt gap is much more complicated. However the qualitative form of the squeezing parameters depicted in Fig.~\ref{fig_3} suggests that the connection between both types of quantities remains valid.

\subsection{Landau-Zener-Stuckelberg approach}
A common theme running through our results is the appearance of large quantum correlations in the regime of intermediate pulse duration in the variation in $\lambda(t)$, or equivalently intermediate ramping velocity. A full many-body theory of this dynamical generation of quantum correlations is not possible at the present, and would likely require a novel theoretical technique for treating Eq.~\ref{Hdicke} in a non-perturbative way. However as a first step towards understanding the complex dynamics discussed here, we consider the simplest version of what happens to a quantum system when it crosses a quantum critical point driven by a time-dependent Hamiltonian. Specifically we provide a heuristic treatment by appealing to the phenomenon of Landau-Zener-Stuckelberg interferometry, by means of which possible trajectories of a quantum system interfere with each other when a transition between energy levels at an avoided crossing (a Landau-Zener transition) is crossed. As discussed in detail in Ref.~\cite{Nori_RMP2014}, when a two-level system is subject to periodic driving with sufficiently large amplitude, a sequence of transitions occurs. The phase accumulated between transitions (commonly known as the Stuckelberg phase) may result in constructive or destructive interference.

Following this heuristic approach, we imagine that we can approximate the complex energy-level diagram of this many-body light-matter system as simply a ground state and a excited-state manifold, separated by some minimum energy gap $\Delta$ during the driving process. During the up-sweep alone, there is a single pass through the avoided crossing (i.e. remnant of the critical point) and so the probability that the system then ends up in this excited state manifold is given by $P_+=P_{\rm LZ}={\rm exp}(-2\pi \Delta^2/4v)$ \cite{Nori_RMP2014}. A similar result follows for the down-sweep alone. However since a pulse involves the double-passage through the avoided crossing region, the resulting probability is given by $P_+=4P_{\rm LZ}(1-P_{\rm LZ}) {\rm sin}^2 \Phi$, where $\Phi$ is the sum of two separate phase contributions: one through the quasi-adiabatic portion and one through the non-adiabatic portion.  Averaging over these phases, and hence averaging over the fine-scale oscillations seen in our results, the probability that the system ends up in the excited state manifold following the pulse is given by ${\overline P_+}=2P_{\rm LZ}(1-P_{\rm LZ})$. As a crude approximate energy scale we set $\Delta=0.5$, which is the value of $\lambda$ at which a purely static QPT occurs in Eq.~\ref{Hdicke}. As $v$ increases, ${\overline P_+}$ rises from zero to a maximum and then decays back to zero. Its maximum value is $0.5$ which corresponds to the maximum entropy scenario in a simple two-level system. We then obtain numerically that ${\overline P_+}$ starts decaying from its maximum when $v\approx 1$. This suggests that the correlation features that we observe should also fall off for $v\rightarrow 1$ ($\Gamma\rightarrow0$), as observed.

\subsection{Future application: system-environment entanglement}
Our findings are also relevant in an entirely different way: if we consider the matter subsystem as the system of interest, and the radiation subsystem as the environment, then our results provide new insight into how a system and its environment become entangled over time, as the system-environment interaction varies. To explore this in the future, instead of considering a single pulse as we do here, the system-environment interaction could be chosen to be a sequence of such pulses which may arrive randomly (e.g. following a Poisson distribution) or become correlated in terms of their arrival times. As such, our model and analysis can provide a first step toward a better understanding of environmental decoherence -- and its flip side, quantum control -- over time. This is important since a primary goal of quantum control is to reliably manipulate quantum systems while preserving advantageous properties such as coherence, entanglement, and purity. Instead of the complex interaction between the system (e.g. matter) and its surroundings (e.g. radiation) being assumed to hamper the system's evolution, it is possible that a suitable sequence of corrective pulses might be used to provide positive feedback to the system and hence maintain its quantum coherences. We leave this for future work.\\

\section{Conclusions} \label{sect_conclu}
We have presented theoretical results for the quantum correlations that develop in a many-body light-matter system, as a result of dynamically manipulating the strength of the light-matter coupling -- specifically, in the form of a single pulse. Our approach was to solve numerically a general, time-dependent many-body Hamiltonian, and exploit the natural partition between radiation and matter degrees of freedom. Specifically, we presented results on intra-subsystem quantum correlations, namely the time-dependent matter and radiation squeezing parameters, and the inter-subsystem Schmidt gap for different pulse duration (i.e. ramping velocity) regimes, from the near adiabatic to the sudden quench limits. The results reveal that both types of quantities signal the emergence of the superradiant state when the quantum critical point is dynamically crossed, by the maximal value of the squeezing parameters and the vanishing of the gap. It is also observed that beyond the near adiabatic limit, the light and matter subsystems remain entangled even when they become uncoupled at the end of the pulse, which could be exploited as a protocol to engineer entangled states of non-interacting systems. Thus our results should also be of interest for temporal control schemes in practical quantum information processing and quantum computation. On a more fundamental level, our results may be helpful for the development of an open-dynamics quantum simulator, for shedding new light on core issues at the foundations of physics, including the quantum-to-classical transition and quantum measurement theory~\cite{Zurek1}, and characterization of Markovianity in quantum systems~\cite{Bruer_PRL, PRBluis,Cosco2017arxiv}. Our findings could also help shed light on system-environment entanglement, if we view the matter subsystem as the system of interest and the radiation subsystem as the environment, and if the system-environment interaction is chosen to be a sequence of pulses with different correlation properties. 

\section*{Acknowledgments}
F.J.G-R., J.J.M-A, F.J.R., and L.Q. acknowledge financial support from Facultad de Ciencias through UniAndes-2015 project \emph{Quantum control of nonequilibrium hybrid systems-Part II}.  F.J.G.R and F.J.R acknowledges financial support from P17.160322.011/01-FISI05 Proyectos Semilla-Facultad de Ciencias at Universidad de los Andes (2017-II).
O.L.A. acknowledges support from NSF-PHY-1521080, JILA-NSF-PFC-1125844, ARO, AFOSR, and MURI-AFOSR. N.F.J. acknowledges support from the National Science Foundation (NSF) under grant CNS 1522693 and the Air Force under AFOSR grant FA9550-16-1-0247. The views and conclusions contained herein are solely those of the authors and do not represent official policies or endorsements by any of the entities named in this paper.
\section*{References}
\bibliographystyle{iopart-num}
\bibliography{\jobname}
\end{document}